\documentclass[11pt,a4paper]{article}
\usepackage{jcappub}
\usepackage{subcaption}
\usepackage{float}
\usepackage{xfrac} 
\usepackage{indentfirst}
\usepackage{graphicx,amssymb}
\usepackage{graphicx}
\usepackage{url}

\newcommand{\aap}{{Astron.\ Astrophys. }}

\newcommand{\apjl}{{Astrophys.\ J.\ Lett. }}

\newcommand{\mnras}{{Mon.\ Not.\ Roy.\ Astron.\ Soc. }}
\newcommand{\nat}{{Nature }}

\def\be{\begin{equation}}
\def\ee{\end{equation}}
\def\ba{\begin{align}}
\def\ea{\end{align}}

\def\ap{\approx}

\def\lsim{\raise0.3ex\hbox{$\;<$\kern-0.75em\raise-1.1ex\hbox{$\sim\;$}}}
\def\gsim{\raise0.3ex\hbox{$\;>$\kern-0.75em\raise-1.1ex\hbox{$\sim\;$}}}

\def\theta{\vartheta}
\def\R{{\cal R}}

\renewcommand{\vec}[1]{\boldsymbol{#1}}

\title{Vela as the Source of Galactic Cosmic Rays above 100\,TeV}

\author[a]{M.~Bouyahiaoui,}
\author[b]{M.~Kachelrie\ss,}
\author[a,c]{D.~V.~Semikoz}

\affiliation[a]{APC, Universit\'e Paris Diderot, CNRS/IN2P3, CEA/IRFU,
Observatoire de Paris, Sorbonne Paris Cit\'e, 119 75205 Paris, France}
\affiliation[b]{Institutt for fysikk, NTNU, Trondheim, Norway}
\affiliation[c]{National Research Nuclear University MEPHI (Moscow Engineering 
Physics Institute), Kashirskoe highway 31, 115409 Moscow, Russia}

\abstract{
  We model the contribution of the nearest  young supernova remannt Vela  to
  the local  cosmic ray flux taking into account both the influence of the
  Local Superbubble and the effect of anisotropic diffusion. The dominant
  contribution of this source in the energy region around the cosmic ray knee
  can naturally explain the observed fluxes of individual groups of nuclei
  and their total flux. Adding the CR flux from a 2--3\,Myr old local CR source
  suggested earlier, the CR spectra in the whole energy range between 200\,GeV
  and the transition to extragalactic CRs are described well by the combined
  fluxes from these two local Galactic sources.
}

\date{\today}


\begin{document}

\maketitle

\section{Introduction}

The measured energy spectrum  of cosmic rays (CRs) extends  smoothly
over more than 11~decades as a nearly featureless power law,
$I(E)\propto E^{-\beta}$. One of
its most prominent features is the knee, a break in the all-particle
energy spectrum at the energy $E_k\simeq 4$\,PeV, which was discovered by
Kulikov  and Khristiansen in the data of the MSU experiment already
in 1958~\cite{1959JETP...35....8K}. The second knee corresponds to a
change in the spectral slope of the all-particle energy spectrum at
$\simeq 5\times 10^{17}$\,eV where the slope hardens by $\Delta\beta\simeq 0.2$.
There is a general consensus that the knee in  the total CR spectrum at 
$E_{\rm k} \simeq 4$\,PeV coincides with a suppression of the primary proton
and/or helium flux, and that the composition becomes increasingly heavier in
the energy range between the knee  and $10^{17}$\,eV~\cite{Aglietta:2004np,Antoni:2005wq,IceCube:2012vv,Apel:2013uni}.

Explanations for the origin of the knee fall in two main categories,
connecting it either with a change in the propagation or the injection
of CRs. In the first case, the knee energy may either corresponds to the
rigidity at which the CR Larmor radius $R_{\rm L}$ is of the order of the
coherence length $l_{\rm c}$ of the turbulent magnetic field in the Galactic
disk~\cite{Giacinti:2014xya,Giacinti:2015hva}. Alternatively, the knee
corresponds to a
transition between the dominance of pitch angle scattering to Hall diffusion
or drift along the regular field~\cite{1993A&A...268..726P,Candia:2002we,Candia:2003dk}.  In both cases,  the energy dependence of the confinement time
changes which in turn induces  a steepening of the CR 
spectrum~\cite{1971CoASP...3..155S,1993A&A...268..726P,Candia:2002we,Candia:2003dk,Giacinti:2014xya,Giacinti:2015hva}.
In the second class of models, the knee
is connected to properties in the injection spectrum of the Galactic
CR sources. For instance, the knee might correspond to the maximal
rigidity to which CRs can be accelerated by the 
population of Galactic CR sources dominating the CR flux below 
PeV~\cite{Stanev:1993tx,Kobayakawa:2000nq,Hillas:2005cs}.
Alternatively, the knee may be caused by a break in the source CR energy
spectrum at this rigidity~\cite{Drury:2003fd,Cardillo:2015zda}.
A variant of this model is the suggestion that the spectrum below the
knee is dominated by a single, nearby source and that the knee correspond to
the maximal energy of this specific source~\cite{Erlykin:1997bs,Erlykin:2000jm}.
All these models lead to a sequence of knees at $ZE_{\rm k}$,
a behaviour first suggested by Peters~\cite{Peters61}. 

In the isotropic diffusion approximation one defines a scalar diffusion
coefficient which depends on energy as $D(E) = D_0 (E/E_0)^{\delta}$.
Measurements of the Boron and Carbon fluxes especially by the AMS-02 experiment
are consistent with Kolmogorov turbulence, i.e.\  $\delta=1/3$,  at
rigidities above $\sim 100$\,GV~\cite{Aguilar:2016vqr}.  
The normalisation $D_0$ is only weakly constrained using measurements
of stable nuclei,  but can be restricted considering the ratio of 
radioactive isotopes as, e.g., $^{10}$Be/$^{9}$Be: Fitting successfully
these ratios requires values of the normalisation constant $D_0$ in the
range $D_0=(3-8)\times 10^{28}$cm$^2$/s at $E_0=10$\,GeV~\cite{1998A&A...337..859P,Evoli:2008dv,Johannesson:2016rlh}.
For typical magnetic field strengths of order $\mu$G and maximal length
scales of fluctuations in the turbulent field of order 10\,pc,
numerically calculated diffusion coefficients are two orders of magnitude
below this value for $D_0$. Since $D$ scales for Kolmogorov turbulence
as $D\propto B^{-1/3}$, the magnetic field strengths $B$ would have to be
scaled down by a factor $10^{-6}$ to obtain agreement between the two
approaches. This discrepancy can be resolved, if the diffusion is sufficiently
anisotropic and the magnetic field contains a non-zero component
perpendicular to the Galactic disk~\cite{Giacinti:2017dgt}. As a result,
the number of sources
contributing to the locally observed flux is reduced by two orders of
magnitude. Thus only few sources  contribute to the local CR flux at energies 
above 200\,GeV.

In the energy range between 200\,GeV and 100\,TeV a 2--3\,Myr old local
supernova (SN) can dominate the local CR flux, as shown in
Refs.~\cite{Kachelriess:2015oua,2015ApJ...809L..23S,Kachelriess:2017yzq}.
A local SN event of the same age was deduced from $^{60}$Fe found in sediments
in the ocean crust of the Earth~\cite{Knie:1999zz,Fitoussi:2007ef,2016Natur.532...69W} and on the Moon~\cite{2016PhRvL.116o1104F}. Such a local SN is able to
to resolve the anomalies which were found recently by CR experiments.
This includes the  energy dependence of the proton to helium ratio, the
breaks in the energy spectrum of primary nuclei at the rigidity 200\,GV, the
positron excess, and the ratio $R\simeq 2 $ of positron to antiproton fluxes,
see Refs.~\cite{Kachelriess:2015oua,2015ApJ...809L..23S,Kachelriess:2017yzq}
for details.

The phase of the CR dipole amplitude is constant between $\simeq 20$\,TeV
and 100\,PeV, except for abrupt flip by $180^\circ$ at $\simeq 200$\,TeV.
Similarly, the dipole amplitude is approximately constant above and below
200\,TeV. This behaviour of the dipole anisotropy suggests that two CR sources
located in the two opposite hemispheres relative
to the local magnetic field line dominate the CR flux below and above this
energy~\cite{Kachelriess:2018kwg}.
We suggest  in this work that Vela, a 11\,kyr old supernova
remnant (SNR)at the distance 270\,pc, is the source dominating the local
CR flux above 200\,TeV. We study the expected CR flux from Vela, which
is connected
with the Solar system by a magnetic field line in models of the
global Galactic magnetic field as, e.g., the Jansson--Farrar
model~\cite{Jansson:2012rt}. If this source would be indeed directly
connected to
the Solar system by a magnetic field line, its flux would however overshoot
the locally measured one by 3 orders of magnitude in case of anisotropic
diffusion. Such an excess is avoided, if one takes into account that the
Earth is located inside the Local Superbubble. We use a simplified model
for the structure of the  magnetic
field inside the Local Superbubble similar to the one of
Refs.~\cite{An17,Andersen:2017yyg}, and follow individual
CR trajectories solving the Lorentz equation. Despite of using a simplified
model for the Local Superbubble we obtain a good description of
the fluxes of individual groups of CR nuclei in the knee region and above.
Adding additionally the CR flux from the 2--3\,Myr old source, the CR spectra
in the
whole energy range between 200\,GeV and the transition to extragalactic
CRs are described well combining the fluxes from only these two
local sources.

\section{Theoretical framework}             
\label{approach}

\subsection{Local Bubble and the geometry of the local magnetic field}

The Sun resides in a low-density region of the interstellar medium (ISM)
called the Local Bubble (LB). The LB extends roughly 200\,pc in the
Galactic plane, and 600\,pc perpendicular to it, with
an inclination of about $20^\circ$~\cite{2003A&A...411..447L}.
Observations and simulations~\cite{Breitschwerdt:1999sv,Schulreich:2017dyt}
show
that the bubble walls are fragmented and twisted. Moreover, outflows away from
the Galactic plane may open up the bubble~\cite{Breitschwerdt:1999sv}.
In view of this complicated geometry, we idealise the LB in our numerical
simulation as follows~\cite{Andersen:2017yyg}: We assume for the magnetic
field profile $\vec B(\vec x)$ parallel to the
Galactic plane $(x,y)$ a cylindrical symmetry, i.e.\ we imply that the changes
as function of the Galactic height $z$ are small compared on the considered
length scales. Then  $\vec B(\vec x) $ is only a function of
$r=\sqrt{x^2+y^2}$. We use as a base radius $R$ of the bubble $R=100$\,pc
and set the wall thickness to $w=3$\,pc. We assume inside the bubble and
the wall a clockwise oriented magnetic field for $y > 0$ and an anticlockwise
one for $y<0$. 
The strength of the regular magnetic field depends only on the radius and is
set to $B_{\rm in}=0.1 \mu G$ inside the bubble, $B_{\rm sh}=12\mu G$ in the
wall, and $B_{\rm out} = 1\mu G$ outside the bubble.
The Sun is assumed to be at the centre of the LB, while Vela is
situated at the distance 270\,pc from the Sun at $y=0$.

We interpolate the transition between different magnetic field regimes by
logistic functions $T(r)$, with a transition width parameter
$w_0 = 2\times 10^{-4}$\,pc. For the inside region $r\le R+w$, we set
\be
T_1= \left[1+ \exp\left(-\,\frac{r-R}{w_0}\right) \right]^{-1} .
\ee
For $y>0$, the regular magnetic field $B_{\rm reg}=(B_{\rm x}^2+B_{\rm y}^2)^{1/2}$
is given by
\begin{align}
B_{\rm x} & = +\left[B_{\rm in} (1-T_1)+B_{\rm sh}T_1\right]\sin(\theta) ,
\\
B_{\rm y} & = -\left[B_{\rm in} (1-T_1)+B_{\rm sh}T_1\right]\cos(\theta) ,
\end{align}
while for $y<0$ it is 
\begin{align}
B_{\rm x} & = -\left[B_{\rm in}(1-T_1)+B_{\rm sh}T_1\right]\sin(\theta) ,
\\
 B_{\rm y} & = +\left[B_{\rm in}(1-T_1)+B_{\rm sh}T_1\right]\cos(\theta) .
\end{align}
For the outside region $r\ge R+w$, we set
\be
T_2 = \left[ 1+ \exp\left(-\frac{r-R-w}{w_0}\right) \right]^{-1}
\ee
and
\begin{align}
B_{\rm x} &= B_{\rm sh}(1-T_2) +B_{\rm out}T_2 ,
\\
B_{\rm y} &= B_{\rm sh}(1-T_2) .
\end{align}
The turbulent magnetic field is taken to be randomly directed with a
strength equal to $B_{\rm reg}/2$. The field modes were distributed between
$L_{\rm min} = 1$\,AU and $L_{ \rm max} = 25$\,pc according to an isotropic
Kolmogorov power spectrum. In the actual simulations, only field modes
above  $L_{\rm min}^\prime = 0.01$\,pc were included.

\subsection{Injection spectrum}             

We use as CR injection spectrum for Vela a broken power law in rigidity with
an exponential cut off at the rigidity $\R_{\rm br}=8\times 10^{15}$\,V,
\be
\frac{dN}{dE} \propto \left\{
    \begin{array}{ll}
        E^{-2.2} , & \mbox{\quad if \quad} E < ZE_{\rm br} \\
        E^{-3.1} \exp(-E/(ZE_{\max})), & \mbox{\quad if \quad} E \ge ZE_{\rm br}.
    \end{array}
    \right.
\ee
The normalisation of the spectra for different groups of CR nuclei will
be fixed such that the propagated fluxes at Earth agree with observations.

The injection spectrum steepens at $\R_{\rm br}=2$\,PV  by $\Delta\beta=0.9$.
Such a steepening is motivated e.g.\ by the analysis of
Ref.~\cite{Drury:2003fd}: Including strong field amplification as suggested
by Bell and Lucek~\cite{2001MNRAS.321..433B,2004MNRAS.353..550B}
into a toy acceleration model, these authors found a
break in the energy spectrum of accelerated protons, coinciding for typical
values SNR parameters with the knee region. The strength $\Delta\beta$
of the steepening depends among others on the injection history, and in
a typical test particle ansatz $\Delta\beta=0.9$ was found.

\subsection{Calculation of the flux}             

In order to compute the flux, we injected $10^{4}$~protons per energy
at the position of Vela and propagated them for 12.000\,yr. We calculated the
CR density $n(E)$ in three regions of interest averaging the CR densities
between 8 to 12\,kyr: Around the source, on the bubble wall, and inside
the bubble. The CR flux $F(E)=c/(4\pi)n(E)$ was then computed from the CR
densities in the considered volumes. For energies below 100\,TeV we deduced
the flux from earlier times and higher energies using the scaling relation
$(E_{\rm low}/E_{\rm high})^{1/3} \approx t_{\rm early}/t_{\rm now}$.

We defined the flux around the source considering the $y$-$z$ plan
centred on the source with a thickness of $\Delta x =  5$\,pc and
$\Delta y =  100$\,pc, and $\Delta z =  100$\,pc form -50\,pc to +50\,pc
on each side.  For the flux inside the bubble wall, we considered a ring
of 1\,pc thickness at the shell, and computed the flux from $ z = -50$\,pc
to $z =50$\,pc. Finally, we computed the flux at the position of the
Earth from the CR density inside a cube of 100\,pc side length centred
at the Sun.

\section{Cosmic ray flux from Vela}            
\label{results}

\begin{figure}[h!]
   \begin{minipage}[b]{1.\linewidth}
      \centering \includegraphics[width=0.66\linewidth]{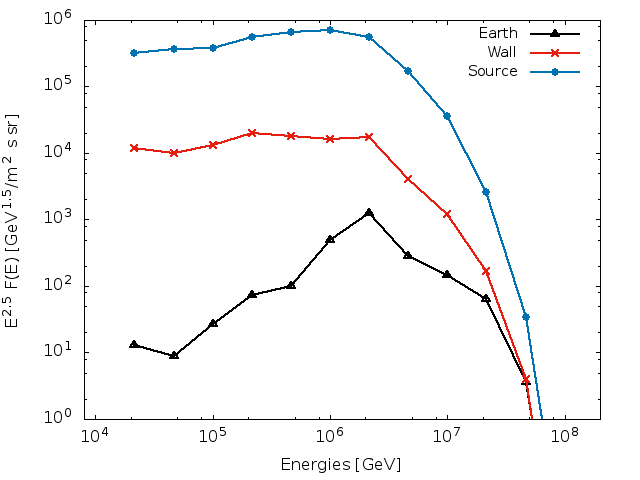}
      \caption{Flux of protons  as function of energy computed inside the bubble centred on the Sun's  position, on the bubble wall  and around the source.
      \label{fig:earth+wall+source}}
   \end{minipage}\hfill
\end{figure}

In Fig.~\ref{fig:earth+wall+source}, we show the normalised proton flux in
the bubble wall, inside the bubble and around the source. We can see that
for high energies ($E_{p} > 10^{17}$\,eV) the bubble is transparent, since
the Larmor radius ($R_L\sim 100$\,pc) of such protons is large compared to
the thickness of the bubble wall. For energies below 1\,PeV, particles start
to be  trapped in the wall and the flux inside the bubble is increasingly
suppressed.

\begin{figure}
\begin{subfigure}{.5\textwidth}
  \centering
  \includegraphics[width=1.\linewidth]{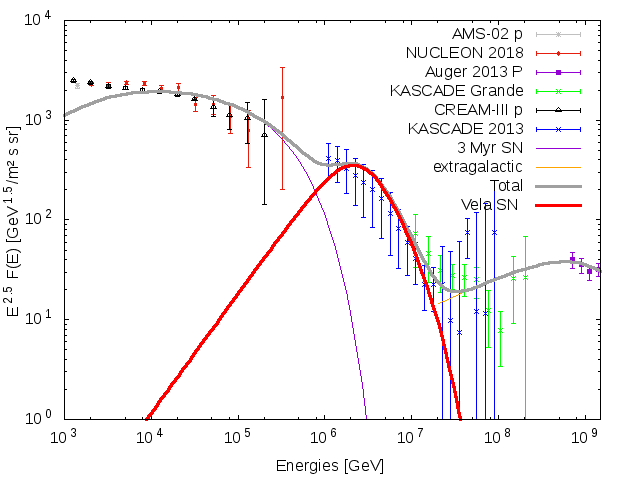}
  \caption{Proton}
  \label{fig:sfig1}
\end{subfigure}%
\begin{subfigure}{.5\textwidth}
  \centering
  \includegraphics[width=1.\linewidth]{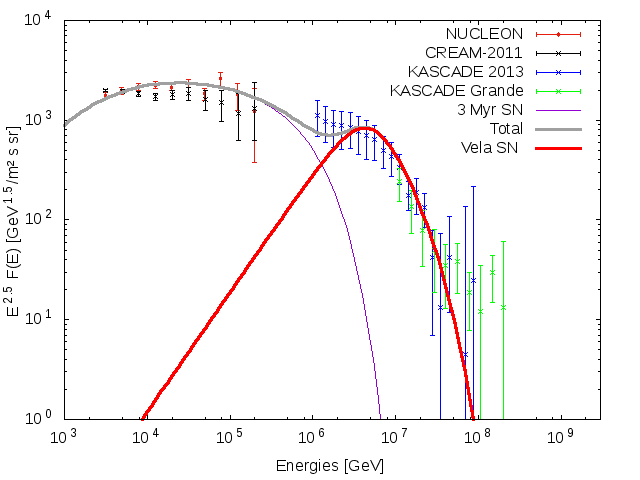}
  \caption{Helium}
  \label{fig:sfig2}
\end{subfigure}
\caption{Proton (left) and Helium (right) flux as
  function of energy   from experiments NUCLEON~\cite{Gorbunov:2018}, CREAM~\cite{Yoon:2011cr}, KASCADE and
  KASCADE-Grande~\cite{Apel:2013uni} and AUGER \cite{Bellido:2017cgf}. 
 Vela flux shown with red line, flux from 2--3 Myr SN  with violet line, extragalactic proton flux with orange line and total flux with black line.
\label{fig:p+he}}
\end{figure}
\begin{figure}
\begin{subfigure}{.5\textwidth}
  \centering
  \includegraphics[width=1.\linewidth]{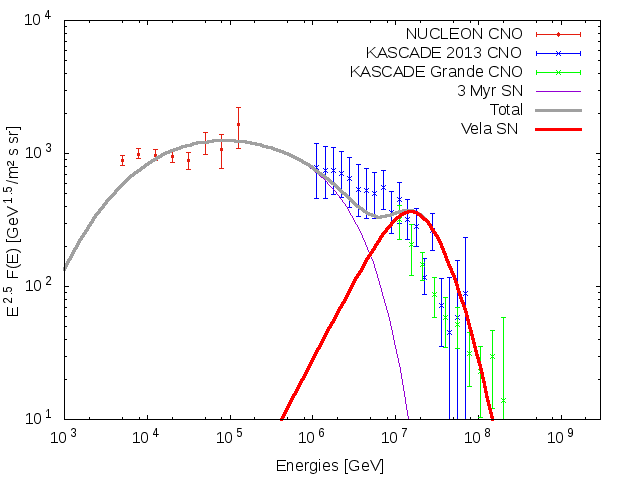}
  \caption{CNO}
  \label{fig:sfig3}
\end{subfigure}
\begin{subfigure}{.5\textwidth}
  \centering
  \includegraphics[width=1.\linewidth]{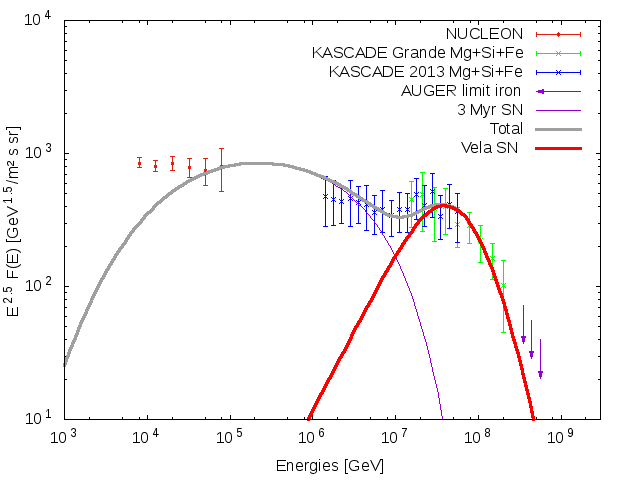}
  \caption{Fe+Si+Mg}
  \label{fig:sfig4}
\end{subfigure}
\caption{Flux for the CNO nuclei (left) and the Mg-Si-Fe group (right)
  from Vela and the 2--3 Myr SN as a function of energy; in the violet
  the Auger limit on the iron flux. Both with same experimental data as
  in Fig.~\ref{fig:p+he}. \label{fig:cno+fesimg}}
\end{figure}
\begin{figure}
\begin{subfigure}{1.\textwidth}
  \centering
  \includegraphics[width=0.66\linewidth]{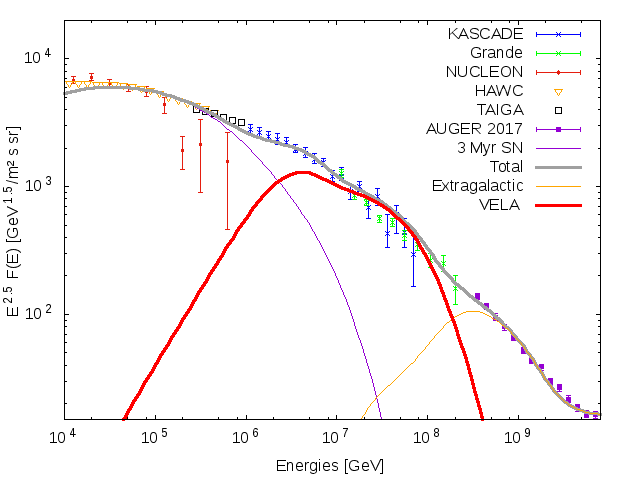}
  \caption{All particles}
  \label{fig:sfig5}
\end{subfigure}
\caption{The all-particles flux from Vela and from the 2--3\,Myr SN  and the
  extragalactic contribution from Ref.~\cite{Kachelriess:2017tvs} together with
  experimental data from NUCLEON~\cite{Gorbunov:2018}, HAWC~\cite{Alfaro:2017cwx}, TAIGA~\cite{Berezhnev:2015zga}, CREAM~\cite{Yoon:2011cr}, KASCADE and KASCADE~Grande \cite{Apel:2013uni}, and AUGER~\cite{Fenu:2017hlc}.
  \label{fig:all}}
\label{fig:fig}
\end{figure}

In Fig.~\ref{fig:sfig1}, we show the proton flux received at Earth
from Vela together
with the proton flux from a 2--3\,Myr old SN in the model of
Refs.~\cite{Kachelriess:2015oua,Kachelriess:2017yzq}. The combined flux of
these two sources covers the energy range from 200\,GeV up to the
extragalactic transition region, fitting well the experimental data.
Additionally, we show the extragalactic proton flux which we obtained from
a fit to the AUGER data as
$$
E^{2.5} F(E) = 5.10^{10}
\left(\frac{E}{10^{16} {\rm eV}}\right)^{0.3}\exp\left(\frac{-E}{1.5 \times 10^{18}{\rm eV}}\right) \,\frac{\rm GeV^{1.5}}{\rm m^2 \, s \, sr} .
$$
We also compute the flux for other nuclei: the flux of helium is shown in
Fig.~\ref{fig:sfig2}, of the CNO group in Fig.~\ref{fig:sfig3} and of the
SiMgFe group in Fig.~\ref{fig:sfig4}, respectively.
From Fig.~\ref{fig:all}, we see
that the all-particles flux fits well the experimental data up to $10^{17}$\,eV.
In the energy range above $10^{17}$\,eV, the extragalactic contribution
becomes important which  we model following
Ref.~\cite{Kachelriess:2017tvs}. We also remark that the iron flux shown
in Fig.~\ref{fig:sfig4} is consistent with the latest Anger composition data
which limit the iron flux to $20 \%$ of the all-particles
flux~\cite{unger:2017cr}. 

We computed the total energy output of Vela from the normalisation of the
simulated data to the experimental ones:  The relative energy fraction
in protons found is~0.3, the one of helium~0.5, of carbon~0.04 and
of iron~0.04, respectively. For the other nuclei, we calculated the flux
ratios of the different nuclei using data from the NUCLEON experiment, taking
for each nuclei three points with same energy and with the smallest error-bars.
Averaging then the ratios, we obtain $F_{\rm N}/F_{\rm C}= 0.25$,
$F_{\rm O}/F_{\rm C}= 1.6$,
$F_{\rm Ne}/F_{\rm C}= 0.3$ ,  $F_{\rm Mg}/F_{\rm Fe}= 0.6$, and
$F_{\rm Si}/F_{\rm Fe}= 0.6$. We obtain then as total energy output in CRs 
$9 \times 10^{49}$\,erg. The total kinetic energy of the Vela supernova
calculated in Ref.~\cite{Sushch:2010je} is $1.4 \times 10^{50}$\,erg.
We note also that
the CR acceleration efficiency of Vela should be high, as it is expected
in the scenario of strong magnetic field amplification of
Refs.~\cite{2001MNRAS.321..433B,2004MNRAS.353..550B}. Since the dependence of
$E_{\rm CR}$ on the exact numerical values of the parameters (like e.g.\
the width and the magnetic field strength of the bubble wall) in our model
is rather large, we conclude that the two values are in good agreement.

\section{Conclusions}

In the standard diffusion picture it is assumed that Galactic CRs form a
smooth, stationary ``sea'' around the Galactic disk. Evidence for this
assumption comes from $\gamma$-ray observations, which indicate a
rather small variation of the parent CR populations below $\ap 100$\,GeV
throughout the Galaxy outside of several kpc from the Galactic
center~\cite{Aharonian:2018rob}. Going to higher energies, CRs escape faster
and thus the number of CR sources contributing to the local flux diminishes.
In order to match the required diffusion coefficient with micro-gauss magnetic
fields observed in the local Galaxy  the CR propagation should be strongly
anisotropic~\cite{Giacinti:2017dgt}. Then the number of CR sources decreases
by a factor~100 relative to the case of isotropic diffusion. As a result,
the CR flux should be dominated by few local CR sources except for the
lowest energies.

In this work, we have examined the suggestion put forward in
Refs.~\cite{Erlykin:1997bs,Erlykin:2000jm} that the spectrum below the
knee is dominated by CRs accelerated in the Vela SNR and that the knee
corresponds to the maximal energy of this source. As an important improvement
compared to these earlier studies, we have taken into account that the Sun
is located inside the Local Superbubble and that CRs propagate anisotropically.
Without the influence of the Local Superbubble, the CR flux from Vela at the
position of the Sun
would overshot the observed one by 3~order of magnitude, because the Sun and
Vela are connected by field lines of the regular magnetic field.
Using a CR injection spectrum with a break $\Delta\beta\simeq 0.9$ at
$E_{\rm br}= 2$\,PeV as motivated by studies of Ref.~\cite{Drury:2003fd}, 
we have obtained a good description of the flux of individual groups of CR 
nuclei both in the knee region and above.
Adding additionally the CR flux from the 2--3\,Myr old source suggested in
Ref.~\cite{Kachelriess:2015oua,2015ApJ...809L..23S,Kachelriess:2017yzq}, the
CR spectra in the
whole energy range between 200\,GV and the transition to extragalactic
CRs are described well combining the fluxes from only these two
Galactic sources.

Finally, we stress that, while including the effect of the Local Superbubble
is an important improvement, the uncertainties connected to the strength and
shape of the magnetic field in the bubble are large. In a future study, we
plan therefore to study in depth the dependence of the spectrum and amplitude
of the CR flux from Vela received on Earth on the parameters and the geometry
of the Local Superbubble. Another
important question to be addressed is how strong the dipole anisotropy from
Vela will be decreased, since the CR flux is effectively emitted not
by a point source but the bubble wall. Last but not least, we note that
the suggestion
from Ref.~\cite{Andersen:2017yyg} that the Galactic soft neutrino
component~\cite{Neronov:2018ibl} in the ICeCube data is produced by CRs
interacting in the wall of a superbubble fits well in the scenario presented
here.

\acknowledgments
\noindent
MK would like to thank APC and Universit\'e Paris Diderot for hospitality
and financial support.



\providecommand{\href}[2]{#2}\begingroup\raggedright\endgroup

\end{document}